\documentclass[%
reprint,
superscriptaddress,
 amsmath,amssymb,
 aps,
floatfix,
]{revtex4-1}

\usepackage[dvipsnames,svgnames,x11names,hyperref]{xcolor}
\usepackage{siunitx}

\usepackage{graphicx}
\usepackage{dcolumn}
\usepackage{bm}
\usepackage{hyperref}
\usepackage[mathlines]{lineno}
\usepackage{mathrsfs}
\usepackage[normalem]{ulem}

\usepackage[margin=1in]{geometry}

\usepackage[normalem]{ulem}
\hypersetup{colorlinks=true, 
	linkcolor={blue!75!black!80!yellow},
	citecolor={blue!75!black!80!yellow}, 
	urlcolor=black 
	}
\makeatletter \renewcommand\@make@capt@title[2]{%
\@ifx@empty\float@link{\@firstofone}{\expandafter\href\expandafter{\float@link}}%
\sffamily{\textbf{#1}}\@caption@fignum@sep#2 }

\begin{document}
\preprint{APS/123-QED}

\title{Nanomagnonic cavities for strong spin-magnon coupling}

\author{Tom\'{a}\v{s} Neuman}
\email{tomasneuman@seas.harvard.edu}
\affiliation{Harvard John A. Paulson School of Engineering and Applied Sciences, Harvard University, Cambridge, MA 02138, USA}

\author{Derek S. Wang}
\affiliation{Harvard John A. Paulson School of Engineering and Applied Sciences, Harvard University, Cambridge, MA 02138, USA}

\author{Prineha Narang}
\email{prineha@seas.harvard.edu}
\affiliation{Harvard John A. Paulson School of Engineering and Applied Sciences, Harvard University, Cambridge, MA 02138, USA}

\begin{abstract}
We present a theoretical approach to use ferro- or ferrimagnetic nanoparticles as microwave nanomagnonic cavities to concentrate microwave magnetic fields into deeply subwavelength volumes $\sim 10^{-13}$\,\si{\milli\meter^3}. We show that the field in such nanocavities can efficiently couple to isolated spin emitters (spin qubits) positioned close to the nanoparticle surface reaching the single magnon-spin strong-coupling regime and mediate efficient long-range quantum state transfer between isolated spin emitters.  Nanomagnonic cavities thus pave the way towards magnon-based quantum networks and magnon-mediated quantum gates.  
\end{abstract}
\date{\today}

\maketitle



\noindent
The study of magnonics, or collective microwave excitations of electronic spins in ferromagnetic and ferrimagnetic materials, has been spurred by advances in superconductor-based quantum systems that have enabled efficient high-fidelity control of quantum states of light and matter \cite{Kjaergaard2019}.
Recent advances have revealed strong coupling between a microwave cavity photon and magnons of a magnetic (nano)particle \cite{zhang2014scmagmw, li2019magnonstrongcoupling, hou2019scnanomagnet}. Interactions of localized magnons among themselves \cite{pirmoradian2018topomagnon} and with microwave cavities \cite{Tabuchi2016quantummagnonics, kong2019magnonnonreciprocity} could be used to engineer nonreciprocal cavity responses \cite{kong2019magnonnonreciprocity} or induce cavity-mediated coupling of magnons to superconducting qubits \cite{wang2020dissipativecoupling},  enabling, for instance, single-shot single-magnon detection \cite{Lachance-Quirion2020singleshotsinglemag}. 
Magnons can also be coherently manipulated and feature strong nonlinearities \cite{elyasi2020nonlinearmagqis}, which makes them attractive for applications in quantum technologies.

Coupling magnons to magnetic emitters \cite{trifunovic2013longdistance, ramos2016, lai2018YIG} is a natural, yet largely unexplored step towards magnon-mediated efficient manipulation of spin-qubit states with applications in sensing \cite{Tribollet2019HybridNS} and quantum information science, for instance, in the field of quantum-state transduction \cite{schuetz2015universaltransducers, rusconi2019SCrings, neuman2020phononic} or design of long-range quantum interconnects \cite{lemonde2018phononnetworks}.
In particular, isolated spins of atoms, molecules \cite{fataftah2018}, nuclear-spin or optically active defects in solids \cite{Childress281, Atature2018, Awschalom2018-en, harris2019group, ciccarino2020strong, Hayee2020-dw} such as the diamond nitrogen-vacancy defect or the silicon-vacancy defect, have become a robust qubit platform maintaining coherence of quantum states for times exceeding seconds \cite{Bradley2019coherence}. Defect-based quantum engineering furthermore relies on interfacing the defects with optical, mechanical or microwave excitations that allow for manipulation of defect states and integration of defect qubits in quantum networks.
Coupling to the fine-structure states of solid-state defects is a challenge that requires the use of spin and orbitally mediated interactions of the defect with its environment \cite{falk2014electrically, golter2016optomechNV, kuzyk2018scaling, lemonde2018phononnetworks,chen2018orbital, maity2018alignment, meesala2018strainsiv, udvarheliy2018spinstrain, li2019honeycomb, Calajo2019quantumacoustooptical, Maity2020straincontrol}. 

We present a scheme using magnetic nanoparticles that sustain antenna-like magnon resonances \cite{lax62, walker1957magnetostatic, walker1958resonant, Fletcher1959ferrimagnetic} as nanomagnonic cavities for microwave magnetic fields. These operate in analogy with optical or infrared metallic nanoantennas \cite{novotny2011antennas, giannini2011plantennas, antennas2015laserphotrev} studied in the field of plasmonics \cite{stockman2018roadmap} that can be used to concentrate optical or infrared electric fields, or silicon nanoparticles able to concentrate magnetic fields in the near-infrared range of the electromagnetic spectrum \cite{Garcia-Etxarri:11, Schmidt:12}. Here, we show that nanomagnonic cavities can modify the local magnetic environment of spin emitters in the microwave domain, facilitate the magnetic drive of spin transitions, and allow for strong coupling of these emitters with single magnons in the single-magnon regime.  
This is possible because these nanomagnonic cavities can concentrate GHz magnetic fields to deeply subwavelength mode volumes down to $\sim 10^{-13}$\,\si{\milli\meter^3} (compared to $\sim 1$\,\si{\milli\meter^3} achievable in transmission-line-based microwave cavities)
and thus enhance the coupling to spin emitters, reaching coupling strengths exceeding intrinsic losses and decoherence of both systems. This single magnon-spin strong coupling regime potentially enables applications including optical single-magnon sensing, magnon-mediated quantum state transduction to spin qubits, or engineering of magnon-mediated nonreciprocal defect-spin networks. 

\begin{figure}
    \centering
    \includegraphics[width=0.9\columnwidth]{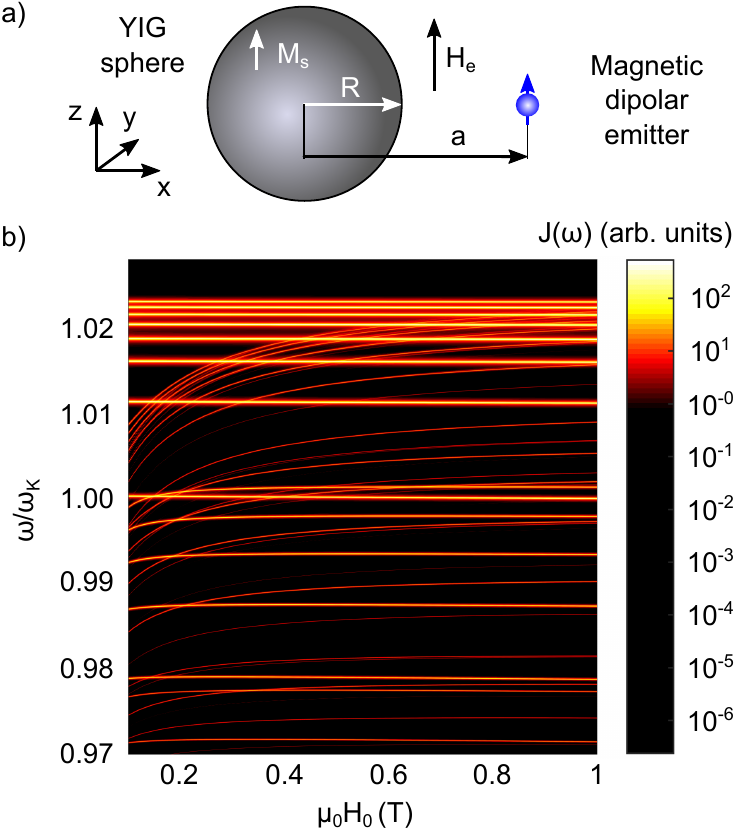}
    \caption{Magnon modes of a sphere interacting with a defect spin transition. (a) A spin emitter is placed at a distance $a$ from the center of a YIG sphere of radius $R$. The sphere is homogeneously magnetized along $z$ to a saturation magnetization ${\bf M}_{\rm s}$. Homogeneous external magnetic field ${\bf H}_{\rm e}$ is applied along $z$. (b) Magnon spectral density $J(\omega)$ for $a=1.2R$, $R=50$\,nm as a function of the magnetic field $H_0$. For presentation purposes we consider spherical multipoles up to $n_{\rm max}=7$, $\Gamma=10/(2\pi)$\,\si{MHz}, and use a nonlinear color scale. The dipolar Kittel mode ($\omega/\omega_{\rm K}=1$) is spectrally surrounded by a large number of higher-order modes whose spectral position relative to $\omega_{\rm K}$ varies as a function of $H_0$.}
    \label{fig:introschematic}
\end{figure}
In the low frequency limit, magnonic responses of ferromagnetic and ferrimagnetic materials can be understood within the macroscopic magnetic response theory \cite{Mills_1974, kit63}. In particular, we assume that the magnetic intensity ${\bf H}$ is related to the magnetic induction ${\bf B}$ via the linear magnetic response ${\bf B}=\mu_0[{\bf I}+\boldsymbol{\chi}(\omega)]\cdot{\bf H}$,
where $\mu_0$ is the vacuum permeability, ${\bf I}$ is the identity tensor, and the magnetic susceptibility tensor $\boldsymbol{\chi}(\omega)$ is defined by ${\bf M}=\boldsymbol{\chi}(\omega)\cdot{\bf H}$.
The magnetic response to external fields can be obtained from the Landau-Lifschitz-Gilbert (LLG) equation \cite{lax62} 
\begin{align}
    \chi_{xx}&=\chi_{yy}=\frac{\gamma^2 H_0 M_{\rm s}}{\gamma^2H_0^2-\omega^2-{\rm i}\Gamma\omega}\equiv\chi,\\
    \chi_{xy}&=\chi_{yx}^*={\rm i}\frac{\gamma \omega H_0 M_{\rm s}}{\gamma^2H_0^2-\omega^2-{\rm i}\Gamma\omega}\equiv{\rm i}\kappa.
\end{align}
Here $\gamma$ is the gyromagnetic ratio $\gamma/(2\pi)\approx 28$\,\si{\giga\hertz\cdot \tesla^{-1}} which we have defined as a positive quantity even for negatively charged electrons. $\Gamma$ is a phenomenological damping parameter related to the Gilbert parameter $\alpha$ as $\Gamma\approx 2\alpha \gamma H_0$, and ${\bf H}_{0}=H_0{\bf e}_z={\bf H}_{\rm e}+{\bf H}_{\rm d}$ (with ${\bf e}_z$ being the $z$-polarized unit vector), where ${\bf H}_{\rm e}$ is the external static field and ${\bf H}_{\rm d}$ is a demagnetization field associated with the shape of the magnetic particle and the particle saturation magnetization ${\bf M}_{\rm s}=M_{\rm s}{\bf e}_z$. For a homogeneously magnetized spherical particle ${\bf H}_{\rm d}=-{\bf M}_{\rm s}/3$ \cite{walker1957magnetostatic, walker1958resonant}. 

To describe the magnetic response of magnetic nanoparticles we invoke the Poisson equation:
\begin{align}
    \nabla\cdot [{\bf I}+\boldsymbol{\chi}(\omega)] \cdot\nabla \phi(\omega)=0, \label{eq:poisson}
\end{align}
where the magnetic scalar potential $\phi$ is linked with the quasi-static magnetic field by ${\bf H}(\omega)=-\nabla\phi(\omega)$. The spatial dependence of $\boldsymbol{\chi}(\omega)$ defines the shape of the magnetic nanoparticle. As a concrete example we consider in the following a magnetic material described by a saturation magnetization of $\mu_0 M_{\rm s}=0.178$\,\si{\tesla}. This closely corresponds to the value observed experimentally in yttrium-iron garnet (YIG) \cite{Wu2010yig}, which is a widely studied ferrimagnetic oxide (ferrite) sustaining long-lived magnonic oscillations (lifetime $> 1$~\si{\micro\second} \cite{kosen2019mwmagnondamping}).


The spin defect is described as a point magnetic emitter of magnetic dipole moment $\hat{\bf m}=-\mu_{\rm B}\hat{\boldsymbol{\sigma}}$, where $\hat{\boldsymbol{\sigma}}$ is the Pauli vector and $\mu_{\rm B}$ is the Bohr magneton. The spin emitter is coupled to the magnon magnetic field by the interaction Hamiltonian $H_{\rm i}=-\mu_0\hat{\bf H}\cdot \hat{\bf m}$ where $\hat{\bf H}$ is the magnetic-field operator.
We note that the spin transition is associated with a circularly polarized transition magnetic dipole moment as we detail in the Supplementary Material.

To study the magnon-spin coupling we place a spin defect close to the surface of a YIG nanosphere as shown in Fig.\,\ref{fig:introschematic}. We then calculate the decay of the excited state of the spin emitter into the magnon modes of the sphere. We obtain the full non-Markovian dynamics of the excited state by invoking the Weisskopf-Wigner approach leading to the integro-differential equation for the excited-state coefficient $c_{\rm e}$ of the spin emitter ($|c_{\rm e}|^2$ is the excited-state population):
\begin{align}
    \dot{\tilde{c}}_{\rm e} = -\int_0^t\int_{-\infty} ^\infty J(\omega)e^{i(\omega_0-\omega)(t-t')}{\rm d}\omega{\tilde{c}}_{\rm e}(t')\,{\rm d}t',\label{eq:IDEdef}
\end{align}
with ${\tilde{c}}_{\rm e}(t')e^{-i\omega_0 t'}={{c}}_{\rm e}(t')$, and
\begin{align}
    J(\omega)&=\frac{\mu_0|\mu_{\rm B}|^2}{\hbar\pi}k_0^2({\rm Im}\left\{ [{\bf G}_{\rm m}]_{xx}+[{\bf G}_{\rm m}]_{yy} \right\}\nonumber\\
    &+{\rm Re}\left\{ [{\bf G}_{\rm m}]_{xy}-[{\bf G}_{\rm m}]_{yx} \right\}),\label{eq:specdenJ}
\end{align}
where ${\bf G}_{\rm m}({\bf r}, {\bf r}')$ is the magnetic Green's tensor that generates magnetic field ${\bf H}_{\rm cl} ({\bf r}) = k_0^2 {\bf G}_{\rm m}({\bf r}, {\bf r}')\cdot {\bf m}_{\rm cl}$ at point ${\bf r}$ induced by a magnetic point dipole ${\bf m}_{\rm cl}$ positioned at ${\bf r}'$ and oscillating at frequency $\omega$, and $k_0=\omega/c$ with $c$ being the speed of light. The Green's tensor can be finally expanded into magnon modes represented by solid harmonic functions \cite{mills2004dimermagnon, walker1957magnetostatic, walker1958resonant, Fletcher1959ferrimagnetic}. Details of the derivation of Eq.\,\eqref{eq:IDEdef}, Eq.\,\eqref{eq:specdenJ}, and the magnetic Green's tensor are shown in the Supplementary Material. 

\begin{figure}
    \centering
    \includegraphics[width=0.95\columnwidth]{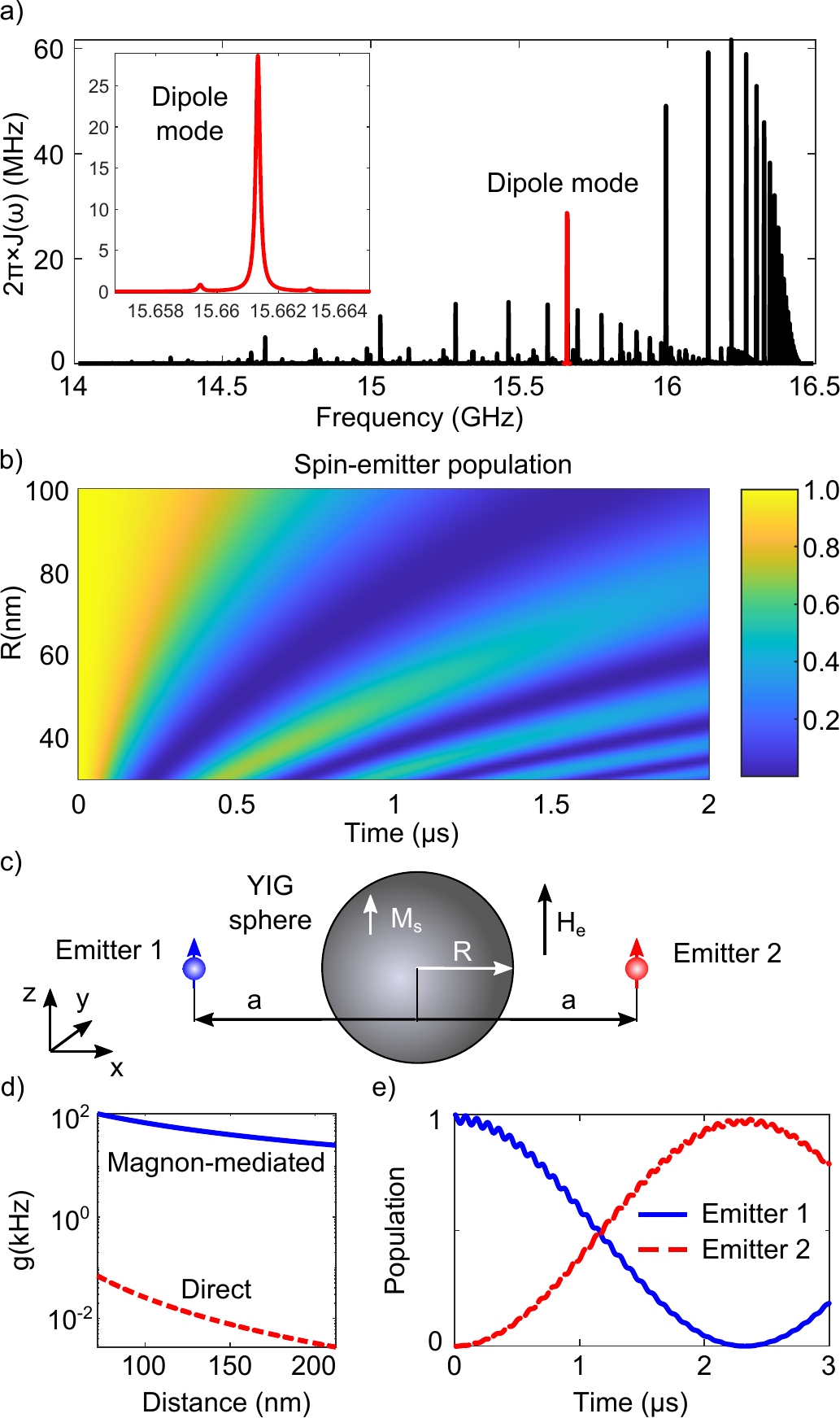}
    \caption{Magnon modes of a sphere interacting with a defect spin transition. (a) Magnon spectral density $J(\omega)$ for $a=1.2R$, $R=30$\,nm, and $\mu_0H_0=0.5$T. The spectral peak corresponding to the dipole surface mode is plotted in red and shown in the inset. Spherical multipoles up to $n_{\rm max}=25$ have been considered. (b) Dynamics of the spin excited state $|c_{\rm e}|^2$ as a function of the sphere radius. The spin transition frequency is tuned to the frequency of the magnon dipole mode $\omega_0=\omega_{\rm K}$. (c) Schematic depiction of two spin emitters coupled via a magnonic excitation. (d) The effective magnon-mediated emitter-emitter coupling (blue solid line) as a function of the emitter-emitter distance $2a$ compared to the direct dipole-dipole coupling between the two emitters separated by the same distance in a vacuum (red dashed line). (e) Dynamics of the populations of two emitters - emitter 1 (blue solid line) and emitter 2 (red dashed line), coupled dispersively by the dipolar magnon mode assuming $R=30$\,\si{\nano\meter} and $a=36$\,\si{\nano\meter}.}
    \label{fig:dynamics}
\end{figure}
The spectral density $J(\omega)$ calculated as a function of the static magnetic field ${\bf H}_0$ is shown in Fig.\,\ref{fig:introschematic}(b) for $a=1.2R$ and $R=30$\,\si{\nano\meter}. 
The spectral density features a large number of narrow spectral peaks whose frequency varies as a function of $H_0$. Each peak corresponds to a Walker mode of the sphere \cite{walker1957magnetostatic, walker1958resonant, Fletcher1959ferrimagnetic}. The peak at $\omega_{\rm K}=\gamma \mu_0\left( H_0+{M_{\rm s}}/{3} \right)$ is particularly significant as it corresponds to the Kittel mode. Due to its dipolar character, this mode can be efficiently excited by an external microwave magnetic field and thus transduce interactions between the microwave photon and spin defects placed in the near-field region of the nanoparticle. The Kittel mode carries a homogeneous circularly polarized magnetization in the particle volume ${\bf M}\propto{\bf e}^{(-)}$ [where ${\bf e}^{(\pm)}=({\bf e}_x\pm{\rm i}{\bf e}_y)/\sqrt{2}$ with ${\bf e}_x$ and ${\bf e}_y$ being the $x$- and $y$-oriented unit vectors].

We next calculate the dynamics of the spin emitter whose frequency is tuned to $\omega_0=\omega_{\rm K}$ and is positioned at $a=1.2R$ for $\mu_0 H=0.5$\,\si{\tesla}. We plot the corresponding spectral density in Fig.\,\ref{fig:dynamics}(a). For these parameters the dipole peak highlighted in Fig.\,\ref{fig:dynamics}(a) in red is spectrally separated from other intense magnon peaks. We calculate the time evolution of the population of the emitter excited state as a function of the sphere radius $R$ ranging between 30 nm and 100 nm and show the result in Fig.\,\ref{fig:dynamics}(b). The dynamics features Rabi oscillations whose frequency decreases with the sphere size. These oscillations are a signature of the strong spin-magnon coupling regime leading to the coherent exchange of energy between the emitter and a single magnon excitation of the dipolar Kittel mode.      

To further analyze the spin-magnon coupling we quantize the magnon field of the Kittel mode using the canonical prescription \cite{sloan2019nickmagnons}:
\begin{align}
    \iiint \mu_0 \tilde{\bf H}^\ast \cdot\frac{\partial(\omega[{\bf I}+\boldsymbol{\chi}])}{\partial \omega}\bigg|_{\omega_{\rm K}}\cdot\tilde{\bf H}\,{\rm d}^3{\bf r}=\hbar\omega_{\rm K}\label{eq:canonical}
\end{align}
and estimate the spin-single-magnon coupling strength $g=-\mu_0 \tilde{\bf H}\cdot {\bf m}_{xy}$ with ${\bf m}_{xy}=-\sqrt{2}\mu_{\rm B}{\bf e}^{(+)}$. In Eq.\,\eqref{eq:canonical} the integral is performed over the whole space and the derivative is evaluated at the resonance frequency of the magnon mode $\omega=\omega_{\rm K}$.
This expression can be integrated for the Kittel mode and allows us to approximate the operator of the magnonic microwave magnetic field as
$\hat{\bf H}=\tilde{\bf H}\, b + \text{H.c.}$, where $b$ is the bosonic annihilation operator, $\tilde{\bf H}={3V\tilde{H}}/({4\pi})\left( 3{{\bf r}\otimes{\bf r}}/{r^5}-{\bf I}/{r^3} \right)\cdot{\bf e}^{(-)}$
with $\tilde{H}=\sqrt{{\hbar\omega_{\rm K}}/{(\mu_0 V_{\rm eff})}}$, $V_{\rm eff}=3V(M_{\rm s}+3H_0)/M_{\rm s}$ is the effective magnon mode volume, and $\otimes$ denotes tensor product. The mode volume is proportional to the physical volume of the particle $V$, and depends on the ratio of the magnetic field $H_0$ to the saturation magnetization $M_{\rm s}$. For a nanosphere of radius $30$\,nm (large with respect to the 1\,\si{\nano\meter^3} YIG unit cell) and $\mu_0H_0=0.5$\,\si{\tesla} we obtain the mode volume $V_{\rm eff}\approx 0.3\times 10^{-13}$\,\si{\milli\meter^3}. Assuming $a=1.2R$ we finally calculate the associated single magnon-spin coupling strength $g/(2\pi)\approx 1$\,\si{\mega\hertz} corresponding to a Rabi frequency $\Omega/(2\pi)=g/\pi\approx 2$\,\si{\mega\hertz} characterising the rate of exchange of energy between the magnon and the spin. This is corroborated by the revival of the spin population seen in Fig.\,\ref{fig:dynamics}(b) at time $t\approx 0.5$\,\si{\micro\second}. 

We show now that the nanomagnonic cavity can be used for quantum state transduction between distant spin emitters. To avoid intrinsic magnon losses that hinder the emitter-emitter state-transfer fidelity, we consider two spin emitters positioned on opposite sides of the magnetic nanosphere, as shown in Fig.\,\ref{fig:dynamics}(c), that are detuned from the Kittel mode. The detuning $\Delta$ is larger than coupling $g$ and damping $\Gamma$ but much smaller than the detuning from other dominant peaks of the spectral density. In this dispersive-coupling regime only a single magnon mode contributes to the dynamics of the spin emitters and mediates their interaction via a virtual-magnon transition, thus mitigating the intrinsic losses of the magnon. The effective magnon-mediated emitter-emitter coupling can then be estimated as $g_{\rm eff}\approx g^2/\Delta$. 
For comparison, we calculate $g_{\rm eff}$ assuming $\Delta=10 g$ as a function of the emitter-emitter separation $2a$ where we increase the sphere radius and keep $a=R+G$, with $G=6$\,\si{\nano\meter}, and plot it in Fig.\,\ref{fig:dynamics}(d) (blue solid line) alongside with the value of the corresponding dipole-dipole coupling $g_{\rm dip}/(2\pi)=\mu_0\mu_{\rm B}^2/[\hbar(2\pi)^2(2a)^3]$ assuming that the two emitters are in a vacuum  (red dashed line). The long-range magnon mediated coupling is approximately three orders of magnitude stronger than the direct coupling and reaches values of $\sim 100$\,\si{\kilo\hertz} for emitter-emitter separation of $72$\,\si{\nano\meter}.

Using this single-mode approximation, we calculate the state-transfer dynamics between the two spin emitters for $R=30$\,\si{\nano\meter} and $a=36$\,\si{\nano\meter} and show the result in Fig.\,\ref{fig:dynamics}(d) (details about the model are provided in the Supplementary Material). We assume that emitter 1 is originally in its excited state and emitter 2 is initiated in the ground state. The population of emitter 1 (blue line) and emitter 2 (red dashed line) undergo periodic exchange of their population (quantum state) with frequency corresponding to the estimated value of $g_{\rm eff}$. These slow oscillations are modulated by fast dynamics visible as ripples in Fig.\,\ref{fig:dynamics}(d) that are due to the direct coupling of the emitters to the magnon. This demonstrates that magnon-mediated spin interactions can indeed mediate long-range high-fidelity state transfer. 

We have shown that magnetic particles can serve as microwave nanocavities able to squeeze the oscillating magnetic fields of wavelength $\lambda$ to mode volumes deeply below $\lambda^3$. A single magnon of such a cavity can strongly couple to isolated magnetic emitters and can induce vacuum Rabi oscillations of the emitter population. Nanomagnonic cavities can also be used to mediate the energy and state transfer between spatially separated magnetic emitters avoiding the magnon losses by dispersive coupling of the emitters. The strength of the single magnon-spin coupling could be further increased by optimizing the geometry of magnonic cavities beyond the spherical shape. For example, magnetic fields could be strongly enhanced in nanometric gaps between isolated magnetic particles \cite{mills2004dimermagnon}. Alternatively, magnonic waveguides relying on propagating magnon waves \cite{damoneschbach1, damoneschbachslab} could be used to couple arrays of spin-emitters and mediate exotic non-reciprocal interactions among them due to the time-reversal symmetry breaking induced by the external magnetic field leading, for instance, to unidirectional propagation of magnon modes. Indeed, the design of magnetic nanostructures for optimal spin-magnon coupling and magnon-mediated spin-spin coupling remains an open question for future studies in the field of cavity nanomagnonics.

\begin{acknowledgments}
This work was partially supported by the Department of Energy `Photonics at Thermodynamic Limits' Energy Frontier Research Center under grant DE-SC0019140 (approaches to strong-coupling) and by the U.S. Department of Energy, Office of Science, Basic Energy Sciences (BES), Materials Sciences and Engineering Division under FWP ERKCK47 `Understanding and Controlling Entangled and Correlated Quantum States in Confined Solid-state
Systems Created via Atomic Scale Manipulation' (spin qubits).
D.S.W. is supported by the Army Research Office MURI (Ab-Initio Solid-State Quantum Materials) grant number W911NF-18-1-0431 and the National Science Foundation Graduate Research Fellowship.
P.N. is a Moore Inventor Fellow through Grant GBMF8048 from the Gordon and Betty Moore Foundation.
\end{acknowledgments}



\newcommand{\noopsort}[1]{} \newcommand{\printfirst}[2]{#1}
  \newcommand{\singleletter}[1]{#1} \newcommand{\switchargs}[2]{#2#1}

\end{document}